\documentclass[11pt]{article}
\usepackage{epsfig,amsmath,amssymb,euscript}
\usepackage[square,comma,sort&compress,numbers]{natbib}
\usepackage{color}
\usepackage{graphicx}
\usepackage{lineno}

\newcommand{\Keywords}[1]{\par\noindent 
{\small{\em Keywords\/}: #1}}

\begin{document}
\title{Multi-model Cross Pollination in Time}

%\author{Leonard A. Smith}
%\altaffiliation[Also at ]{Pembroke College. University of Oxford. UK}
%\homepage{http://www.lsecats.org}
%\email{lenny@maths.ox.ac.uk}
%\author{Hailiang Du}
%\email{h.l.du@lse.ac.uk}
%\affiliation{Centre for the Analysis of Time Series.\\ Department of
%  Statistics. London School of Economics, London WC2A 2AE. UK
%}

\author{Hailiang Du$^{1,2}$ \quad \quad  Leonard A. Smith$^{1,2,3}$ \\[2ex]
        $^1$Center for Robust Decision Making on Climate and Energy Policy,\\
   University of Chicago, Chicago, IL, US\\[1ex]        
   $^2$Centre for the Analysis of Time Series,\\
       London School of Economics, London WC2A 2AE. UK\\[1ex]
   $^3$Pembroke College, Oxford, UK
   }

\date{\today}

\maketitle

\begin{abstract}

Predictive skill of complex models is often not uniform in model-state space; in weather forecasting models, for example, the skill of the model can be greater in populated regions of interest than in ``remote" regions of the globe. Given a collection of models, a multi-model forecast system using the cross pollination in time approach can be generalised to take advantage of instances where some models produce systematically more accurate forecast of some components of the model-state. This generalisation is stated and then successfully demonstrated in a moderate ($\sim40$) dimensional nonlinear dynamical system suggested by Lorenz. In this demonstration four imperfect models, each with similar global forecast skill, are used. Future applications in weather forecasting and in economic forecasting are discussed. The demonstration establishes that cross pollinating forecast trajectories to enrich the collection of simulations upon which the forecast is built can yield a new forecast system with significantly more skills than the original multi-model forecast system. %in the medium range (that is, $40\%$ more probability at the outcome, on average).

\end{abstract}

\section{Introduction}

Nonlinear dynamical systems are frequently used to model physical processes such as fluid dynamics and weather. Uncertainty in the observations makes identification of the exact state impossible for a chaotic nonlinear system, which calls for forecast based on an ensemble of initial conditions to reflect the inescapable uncertainty in the observations by capturing the sensitivity of each particular forecast. In general, when forecasting real systems, for example the Earth's atmosphere as in weather forecasting, there is no reason to believe that a perfect model exists. Generally the model class from which the particular model equations are drawn does not contain a process that is able to generate trajectories consistent with the data. In order to take into account both the structural model error and uncertainties in initial conditions, the multimodel and ensemble techniques can be combined to a new approach, known as the multi-model ensemble concept (see \cite{Harrison95,Palmer04}). In recent years, multi-model ensembles have become popular tools to investigate and account for shortcomings due to structural model error in weather and climate simulation-based predictions on time scales from days to seasons and centuries (\cite{Kirtman13,Palmer04,Wang09,Weisheimer09}). While there have been some results suggesting that the multi-model ensemble forecasts outperform the single model forecasts in an RMS sense,  {for example \cite{Weisheimer09,Kirtman13}, Smith, et al.~\cite{Smith14} challenged the claim that the multi-model ensemble provides a ``better" probabilistic forecast than the best single model}. The current multi-model ensemble forecasts are based on combining single model ensemble forecasts only by means of statistically processing model simulations to form the forecasts. To the extent that each model is developed independently, every single model is likely to contain different local (regional, for example) dynamical information from that of other model; such information is not expected to be explored by statistical processing. Using statistical processing, such information is only carried by the simulations under a single model ensemble: no advantage is taken to influence simulations under the other models. This paper presents a novel methodology, named Multi-model Cross Pollination in Time, for multi-model ensemble scheme with the aim of integrating the dynamical information from each individual model operationally in time instead of statistical processing. The proposed approach generates model states in time via applying data assimilation scheme(s) over the multi-model forecasts. Illustrated here using the moderate-order Lorenz model~\cite{Lorenz96}, the proposed approach is demonstrated to {allow significant improvement upon} the traditional statistical processing and best single model ensemble. It is suggested this illustration could form the basis for more general results which in turn could potentially be deployed in operational forecasting. In weather forecasting, there is a tendency to focus on model performance locally, North America for National Centers for Environmental Prediction (NCEP), Europe for European Centre for Medium-Range Weather Forecasts (ECMWF) and Asia for Japan Meteorological Agency (JMA).  

The multi-model ensemble forecast problem of interest is defined and traditional statistical processing approaches are reviewed in Section 2. A full review of simple Multi-model Cross Pollination in Time (CPT I) approach is presented in Section 3. An advanced Multi-model Cross Pollination in Time (CPT II) approach is presented in Section 4. 
%The Pseudo-orbit Data Assimilation (PDA) approach~\cite{Du2014a,Du2014b} adopted in CPT II is described in Section 5. 
The experiment based on Lorenz 96 system-models pair is designed as well as the results are presented in Section 5. Section 6 provides discussion for wider applications and conclusions.

%Multi-model outputs are firstly combined in the observation space. The combined outputs are then used to obtain new forecast ensemble of each single model via an advanced data assimilation algorithm so that the obtained model states are dynamically consistent each corresponding model. %Iterating those initial conditions forward under the each model provides a multi-model ensemble forecast in real time. 
%This proposed approach explores the dynamical information of each model without requiring all the models share the same model space.

%This project will demonstrate viability of the proposed approach in low order system and models as well as high order operational models. The evaluation will be provided based on comparing the proposed approach with the state-of-art multi-model ensemble approaches in order to demonstrate significance.

\section{Problem description}

%$\textbf{s}_{t}=h(g(\tilde{\textbf{x}}_{t}))+\boldsymbol{\eta}_{t}$ where $\textbf{s}_{t}\in \mathbb{O}$ and $\boldsymbol{\eta}_{t}$ represents the observational noise

Outside those problems defined within pure mathematics, there is arguably no perfect model for problems including a physical dynamical system~\cite{Smith02,Judd04} evolving smoothly in time. One hypotheses a nonlinear system with state space $\mathbb{R}^{\tilde{m}}$, the evolution operator of the system is $\tilde{F}$ (i.e. $\tilde{\textbf{x}}(t+1)=\tilde{F}(\tilde{\textbf{x}}(t))$ where $\tilde{\textbf{x}}(t)\in \mathbb{R}^{\tilde{m}}$ is the state of the system). $\tilde{F}$, $\tilde{\textbf{x}}$, and $\tilde{m}$ are unknown. It is often useful to speak as if such a system existed, regardless of whether or not one actually does exist. An observation of the system state at time $t$ is defined by $\textbf{s}(t)=\tilde{h}(\tilde{\textbf{x}}(t))+\boldsymbol{\eta}(t)$ where $\textbf{s}(t)\in \mathbb{O}$, $\tilde{h}$ is the observation operator that projects the system state onto observation space and $\boldsymbol{\eta}(t)$ represents the observational noise. What is in hand are $M$ models each of which approximates the system, with the form $\textbf{x}(t+1;i) = F_{i}(\textbf{x}(t;i)), i=1,\dots,M$,  where $\textbf{x}(t;i) \in \mathbb{R}^{m(i)}$, $\mathbb{R}^{m(i)}$ is the model-state space corresponding to the $i^{th}$ model $F_i$. In practice, model-state space usually differs from observation space, and it is likely that different models define different model-state space. The model states can be projected into the observation space via an observation operator $h_{i}(\cdot)$, (different model may also have different operator). 

The simplest reaction to have $M$ models, each of which provides N-member ensemble forecast, might be to identify the best, discard others. If the models are of comparable quality\footnote{or even in the case some models are inferior on average but competing on occasion.}, then it is likely that different models will tend to do better in different regions of state space (weather models for example, on different geographical locations or different synoptic conditions), due to variations in the particular processes that are important locally.\footnote{In practice, there is rarely enough data to identify which one will be the best on a given data~\cite{Higgins16}, and a reasonable alternative is to compute $M$, N-member ensembles, one ensemble under each model and treat each ensemble equally.} 

{Consider each model producing an N-member ensemble forecasts by iterating an N-member initial condition ensemble forward. In practice, such multi-model forecast system is verified using the future observations. The goal here in this paper is to introduce new multi-model ensemble forecast system (in time) to improve\footnote{The improvement is quantified by the information in probabilistic forecasts reflected in $-\log_{2}(p(Y))$ (see~\cite{Good52} and Section 5).} forecast of the future states.

%The superscript $i$ denotes the $i^{th}$ model and the subscript $j$ denotes the $j^{th}$ ensemble member. 
%^{i} X_{0}\equiv

The Model Output Statistics (MOS) has a long and successful history of statistically processing single model ensemble forecasts (see~\cite{Wilks06,Wilks07} and references therein). For multi-model ensemble, statistical approaches have been proposed to combine ensembles of individual model runs to produce a single probabilistic multi-model forecast distribution, mostly based on weighting the models according to some measure of past performance, for example~\cite{Hagedorn05,Rajagopalan02,Doblas05}. The output of these statistical processing approaches is a function of each individual forecast ensemble. Only the single model ensemble forecast are conducted in time and carries some information of the model dynamics. Despite the multi-model ensemble theme is designed to account for model inadequacy as different models have different model structure, statistically processing the models output can hardly explore the local dynamical information of each individual model. The extension of CPT approaches presented in this paper integrate the dynamical information from each individual model in time and produce, truly new, multi-model trajectories which significantly increases the information in the ensemble of simulations beyond that available to the original multi-model ensemble forecast. {This addition is gained by allowing communication between different models regarding trajectories in the future.}

\section{Multi-model Cross Pollination in Time I}

When the different models have independent structural shortcomings, then cross-pollinating trajectories between models to obtain truly multi-model trajectories can allow the ensemble of trajectories to explore important regions of state space the individual models just can't reach. 

Smith~\cite{Smith00} introduced the Cross Pollination in Time (hereafter CPT I) approach exploiting the assumption that all the models share the same model-state space\footnote{or there are known one-to-one maps which link their individual state space, given all the models are iterated discretely.}.
Let $\Delta t$ be the observation time where every $\Delta t$ time step an observation is recorded. For simplicity, at every observation time all the models provide their model outputs\footnote{Note it is often the case that the model iteration(simulation) step is much smaller than the observation time, different models may have different iteration step and produce outputs with different time frequency.}. Let $\tau$ be the cross pollination time that every $\tau$ a cross pollination is taken place. Given $M$ N-member ensembles of trajectories made under each model, firstly consider the ensembles of states at $t=\tau$ as one large ensemble of $N \times M$ states in a model-state space. Secondly using some pruning scheme to reduce this large ensemble to N-member states in order to maintain a manageable ensemble size. While the optimal pruning scheme is still an object of research, the simple approach~\cite{Smith00} of identifying nearest pair of states, and then deleting the one member from this pair of two states with the smallest second nearest neighbor distance, has been found to more effective than random selection in some simple examples.\footnote{Note that the aim pruning is quite different than that of resampling from an estimated PDF~\cite{Bishop01}.} In this paper, a pruning scheme based on the local forecast performance (see Section 5) is adopted to serve the purpose of demonstrating the use of proposed CPT II approach. Thirdly use these $N$ states as initial conditions and propagate them forward under each of the M models to produce $M$ N-member ensembles of trajectory segments until the next cross pollination time $2\tau$. Repeat these three steps until the forecast time of interest is reached, and then interpret the ensemble~\cite{Brocker08}.

Inasmuch as the CPT I ensemble scheme contains implicitly all trajectories of each of its constituent models, the dynamical information of each individual model is explored and integrated. In practice, however, the assumptions of this approach are less likely to hold: different models usually define different model-state space, and the one-to-one maps, which link different model-state space, may not exist for example in weather forecasting. {More relevant for the work below, however, is that CPT I traditionally considers the entire model state, without regarding for the fact that some models might forecast some components with greater skill. Each trajectory segment under CPT I is a trajectory of one of the $M$ models, the cross pollination is of trajectory segments; CPT II aims to use the information in these model dynamics more effectively. Another shortcoming of CPT I is that for each model the initial conditions produced by other models are unlikely to be consistent with that model's dynamics or be efficient and quality samples of initial conditions for that model as iterating those initial conditions under the model for a short period like $\tau$ may lead them to the model attractor.} The CPT II approach introduced in the next section frees one from the assumptions and overcomes such shortcomings.

%These $N$ conditions are then propagated forward under each of the M models. And so on. This approach assumes that either all the models share the same model-state space, or the one-to-one maps exist which link their individual state space; neither needs be the case in practice for example weather forecasting. 

\section{Multi-model Cross Pollination in Time II}

The Multi-model Cross Pollination in Time (CPT II) approach presented here not only frees one from the assumption that all models share the same model-state space but also extracts and integrates the dynamical information from each model via exploring the sequence space.

The CPT II approach consists of three steps:

\begin{itemize}

\item[(i)] Combine Multi-model outputs in the observation space to create an ensemble of orbits, each orbit consist of a sequence of states in the observation space. 

For each individual model, the forecast ensemble is obtained via iterating an initial condition ensemble forward until the first CPT time $\tau$, which produces an ensemble of model trajectory segments, from time $t_{0}$ to $t_{0}+\tau$. Although different models may define different model-state space, every model state can be projected onto observation space using the corresponding observation operator. A model trajectory segment of the $i^{th}$ model, projected onto observation space, becomes an orbit, $\textbf{X}(i)\equiv\{h_{i}((\textbf{x}(t_{0};i)),h_{i}((\textbf{x}(t_{0}+\Delta t;i)),\dots,h_{i}((\textbf{x}(t_{0}+\tau;i)) \}$, where $t_{0}$ is the initial time and $\textbf{x}(t+\Delta t;i)=F_{i}^{\Delta t}(\textbf{x}(t;i))$. For $M$ models and each produces N-member ensemble, it forms one large ensemble of $M \times N$ orbits $\textbf{X}(i,j), i=1,\dots,M$ and $j=1,\dots,N$ in the observation space. There are various statistical processing approaches to combine the multi-model ensemble of sequence states in the observation space, for example traditional MOS approach. And in order to maintain a manageable ensemble size, one may prune this large ensemble back into N sequences of states using some pruning scheme (the pruning scheme used in this paper is described in the following section), that is: 
\begin{eqnarray}
%\{\textbf{X}(i,j)\mid i=1,\dots,M \quad and \quad j=1,\dots,N \} 
\left.
                \begin{array}{ll}
                  \{\textbf{X}(1,1), \dots, \textbf{X}(1,N)\} \\
                  \dots\\
                  \{\textbf{X}(M,1), \dots, \textbf{X}(M,N)\}
                \end{array}
              \right\} \rightarrow \{\textbf{Y}(1),\dots,\textbf{Y}(N)\}
\end{eqnarray}
$\textbf{Y}$ is the combined output of ensemble sequence states in the observation space, $\textbf{Y}(j)\equiv\{ \textbf{y}(t_{0};j),\textbf{y}(t_{0}+\Delta t;j),\dots,\textbf{y}(t_{0}+\tau;j) \}$ where $\textbf{y}(t;j)\in \mathbb{O}$.

\item[(ii)] Data assimilation of locally preferred ensemble signals

Given N sequences of states in the observation space, \{\textbf{Y}(1),\dots,\textbf{Y}(N)\}, each individual model can apply a data assimilation scheme to each sequence of state to obtain a sequence of model states in its model-state space, this corresponds to treat a sequence of states in the observation space as a sequence of observations:
\begin{eqnarray}
 \{\textbf{Y}(1),\dots,\textbf{Y}(N)\} \rightarrow \left\{
                \begin{array}{ll}
                  \{\textbf{Z}(1,1), \dots, \textbf{Z}(1,N)\} \\
                  \dots\\
                  \{\textbf{Z}(M,1), \dots, \textbf{Z}(M,N)\}
                \end{array}
              \right.
\end{eqnarray}
$\textbf{Z}(i,j)$ is the $j^{th}$ sequence of model states in the $i^{th}$ model state space, $\textbf{Z}(i,j)\equiv\{ \textbf{z}(t_{0};i;j),\textbf{z}(t_{0}+\Delta t;i;j),\dots,\textbf{z}(t_{0}+\tau;i;j)  \}$ where $\textbf{z}(t;i;j)\in \mathbb{R}^{m(i)}$.

It is not necessary for each model to apply the same data assimilation scheme (in practice, it is likely that each model has its own data assimilation scheme), using existing data assimilation schemes would clearly avoid extra cost of implementing the CPT II approach. It is, however, noted that applying data assimilation here is crucial in order to extract dynamical information from the model. It is desirable to use nonlinear data assimilation scheme which accounts for structural model error, for example Pseudo-orbit Data Assimilation (see Du and Smith~\cite{Du2014a,Du2014b}), a brief description is given in the Appendix A. As the model is not perfect, the data assimilation scheme used here is not aiming to obtain model trajectories, but pseudo-orbits~\cite{Du2014b}. Projecting the end component of the model pseudo-orbits, obtained from the data assimilation, into the observation space would provide $N\times M$ forecast states at $t=t_0+\tau$.

\item[(iii)] Iterate new model states (from ii) forward

Consider the end component of $\textbf{Z}(i,j)$, $\textbf{z}(t_{0}+\tau;i;j)$, as initial condition for the $i^{th}$ model and $j^{th}$ member and iterate forward using the $i^{th}$ model until the next cross pollination time $t_{0}+2\tau$, which produces an ensemble of model trajectory segments, from time $t_{0}+\tau$ to $t_{0}+2\tau$. 

\end{itemize}

Repeat (i),(ii) and (iii) to provide forecast states at $t=t_{0}+2\tau$ and so on to provide forecasts at $t=t_{0}+k\tau$, $k=3,4,...$.

{Cross Pollination in Time~\cite{Smith00} differs fundamentally from other ``forecast assimilation" techniques. Stephenson et al.~\cite{Stephenson05} for example introduced a novel approach to forecast assimilation which generalizes earlier calibration methods including model output statistics (see~\cite{Wilks05}). This approach provides a map from the space of model simulations to the space of observations. In general any map from the model-state space to the target observable space which uses (only) information available at the time the forecast simulations were launch is admissible.  Brocker and Smith~\cite{Brocker08} discuss other approaches to ensemble interpretations. None of these papers, however, enable the feedback of forecast-simulation information into the dynamics of the forecast itself. CPT II does precisely this.}

\section{Experiments based on Lorenz96}

A system of nonlinear ordinary differential equations (Lorenz96 System) was introduced by Lorenz~\cite{Lorenz96}. For the system containing $n$ variables $x_{1},...,x_{n}$ with cyclic boundary conditions (where $x_{n+1}=x_{1}$), the equations are
\begin{eqnarray}
 \label{eq:Lorenz96 Model I}
    \frac{d x_{i}}{dt}=-x_{i-2} x_{i-1}+ x_{i-1} x_{i+1}-x_{i}+F,
\end{eqnarray}
The system is supposed to represent a one-dimensional atmosphere; the $n$ variables $x_{1},...,x_{n}$ are to be identified with the values of some unspecified scalar atmospheric quantity at $n$ equally spaced points about a latitude circle which is called grid points, even though the ``grid" is one-dimensional.~\cite{Lorenz96} $F$ is a positive constant, it is also found~\cite{Lorenz96} that as long as $n > 12$, chaos is found when $F > 5$.

The {\it true} system (hereafter, system) used in the following experiments, contains $40$ variables, $n=40$, and the values of the parameter $F$ varies with locations, i.e. $F=8$ for $i=1,\dots,10$;  $F=12$ for $i=11,\dots,20$; $F=14$ for $i=21,\dots,30$; $F=10$ for $i=31,\dots,40$. Four models each defined using the same dynamical equation as the system but with fixed value of parameter $F$, that is: model I, $F=8$ for all $i$; model II, $F=12$; model III, $F=14$ and model IV, $F=10$. 

Both the system and the model are defined using a standard fourth-order Runge-Kutta numerical simulation. The simulation time step is $0.01$ time unit and the model time step $\Delta t$ is $0.05$, that is each model time step is conducted by 5 steps of the fourth-order Runge-Kutta integrator. Observations $\textbf{s}(t)\in \mathbb{R}^{40}$ are generated by the system plus IID Gaussian noise,  $N(0,\sigma_{Noise}^2)$, $\sigma_{Noise}=0.2$, at every system time step ($0.05$ time unit).\footnote{In this setting, the model state space, system state space and the observation space are identical, although the proposed CPT II is not constrained to operate in this ideal setting.} The cross pollination time $\tau=0.4$ indicates 2 days\footnote{Assuming $1$ time unit is equal to 5 days, the doubling time of the Lorenz96 system roughly matches the characteristic time-scale of dissipation in the atmosphere (see Lorenz~\cite{Lorenz96}).}.

For a forecast initial time $t_{0}$, a simple inverse noise method (adding draws from the inverse of the observational noise to the observation) is adopted to generate a 9-member initial condition ensemble for all the models, $\textbf{IC}(t_{0})\equiv\{ \textbf{x}(t_{0},j)\in\mathbb{R}^{40}, j=1,...,9\}$. Iterate the initial condition ensemble forward under each of the four models until time $t_{0}+\tau$. For each model this gives an ensemble of model trajectories, for example for model I: $\{\textbf{X}(I,1),\dots,\textbf{X}(I,9)\}$ where $\textbf{X}(I,j)\equiv \{\textbf{x}(t_{0};I;j),\textbf{x}(t_{0}+\Delta t;I;j),\dots,\textbf{x}(t_{0}+\tau;I;j)\}$, $\textbf{x}(t+\Delta t;I;j)=F_{I}^{\Delta t}(\textbf{x}(t;I;j))$ and $\textbf{x}(t;I;j)\equiv\{ x_{1}(t;I,j),x_{2}(t;I,j),\dots,x_{40}(t;I,j)\} \in \mathbb{R}^{40}$ .

For each model, a large set of forecasts is collected by repeating the above experiments for 4096 different initial states. Consider half of the set as training set to fit parameter values and the other half as test set for evaluation. To assess each model's forecasts at various lead time, the forecast ensemble is translated into predictive distribution function by kernel dressing and blending with climatological distribution (for a full description see~\cite{Brocker08}, and Appendix B). 

The forecast performance is evaluated with IJ Good’s logarithmic score (Ignorance score;~\cite{Good52,Roulston02}). The Ignorance score is the only proper local score for continuous variables~\cite{Bernardo79,Raftery05,Brocker07}. Although there are other nonlocal proper scores, the authors prefer using ignorance as in addition it has a clear interpretation in terms of information theory which can be easily communicated in terms of effective interest returns~\cite{Hagedorn09}.\footnote{There are no compelling examples in favor of the general use of nonlocal scores and some nonlocal scores have been shown to produce counter intuitive evaluations~\cite{Smith15}.} Ignorance is defined by:
\begin{eqnarray}
 \label{eq:IGN}
    S(p(y),Y)=-\log_{2}(p(Y)),
\end{eqnarray}
where $Y$ is the outcome and $p(Y)$ is the probability of the outcome $Y$. In practice, given $K$ forecast-outcome pairs $\{(p_{i},Y_{i})\mid i=1,\dots,K\}$, the empirical average Ignorance score of a forecast system is then  
\begin{eqnarray}
 \label{eq:eIGN}
    S_{E}(p(y),Y)=\frac{1}{K}\sum_{i=1}^{K}-\log_{2}(p_{i}(Y_{i})),
\end{eqnarray}

\begin{figure}[!h]
\hbox{
  \epsfig{file=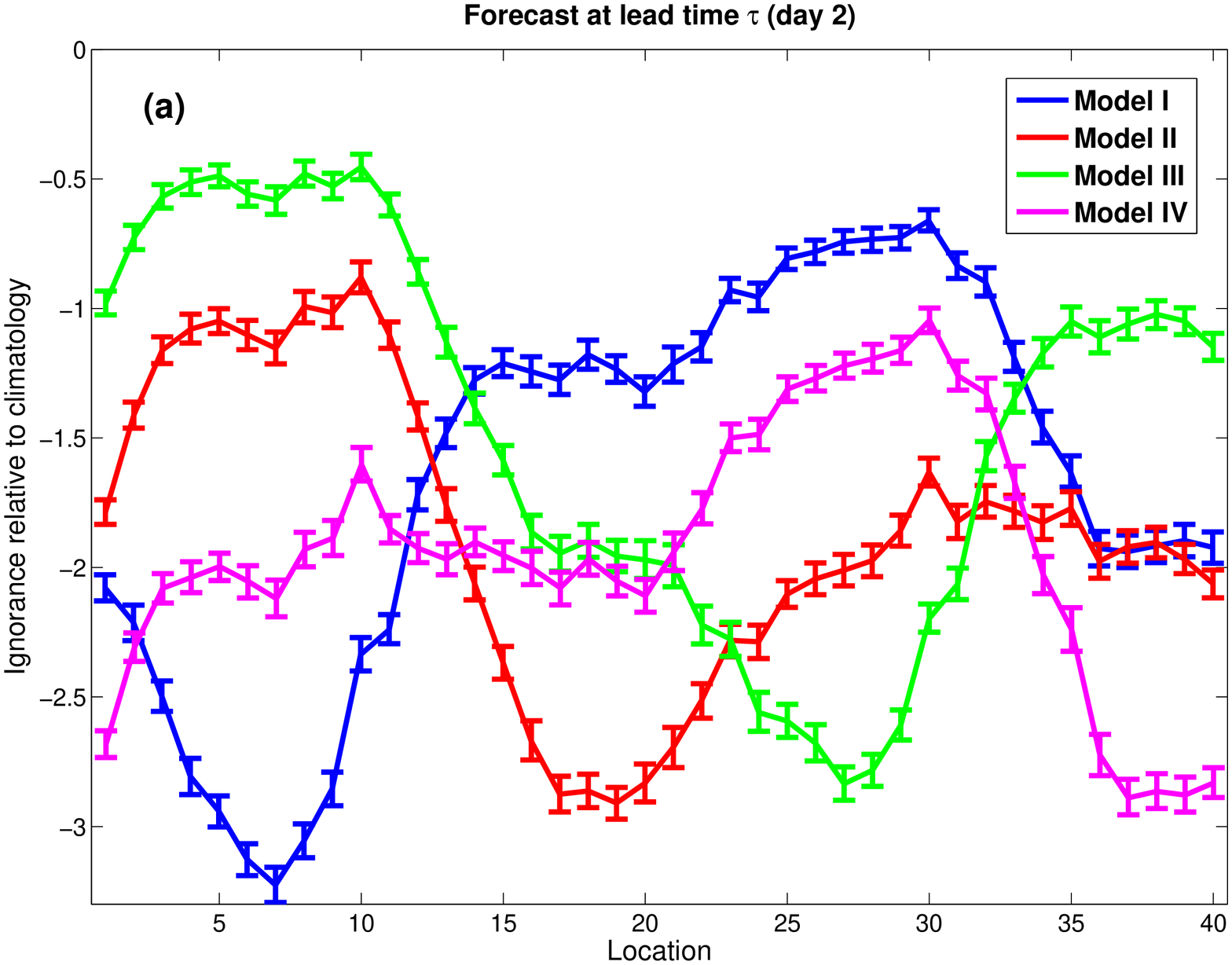, width=0.48\columnwidth, height=6cm}
  \epsfig{file=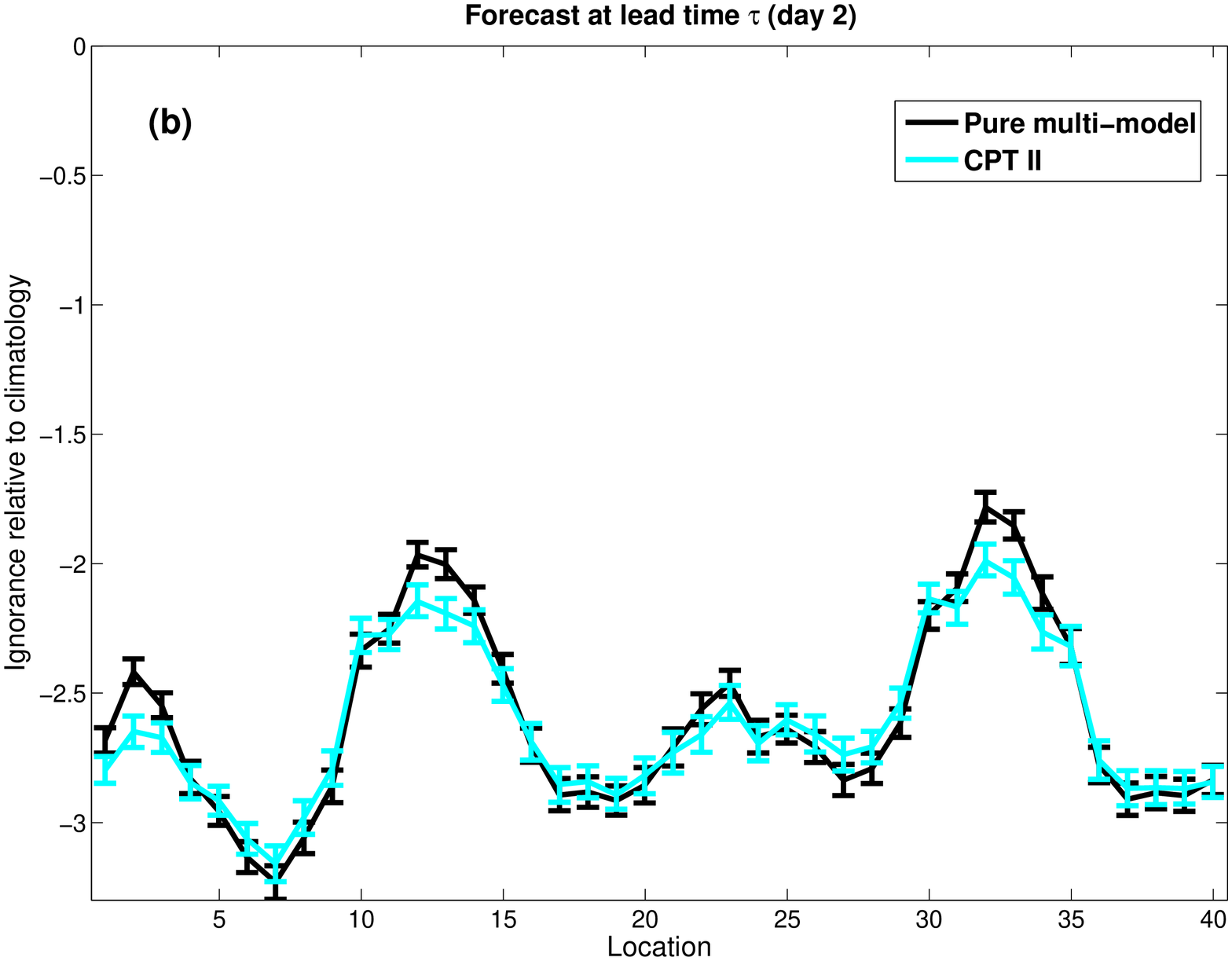, width=0.48\columnwidth, height=6cm}
}

\caption{Ignorance score of forecasts as a function of location (dimension) at lead time $\tau=0.4$ time unit, a) forecasts from each individual model, b) pure multi-model forecast (Black) and CPT II forecast (Cyan).}
\label{fig:maperr1}
\end{figure}

\begin{figure}[!h]
\hbox{
  \epsfig{file=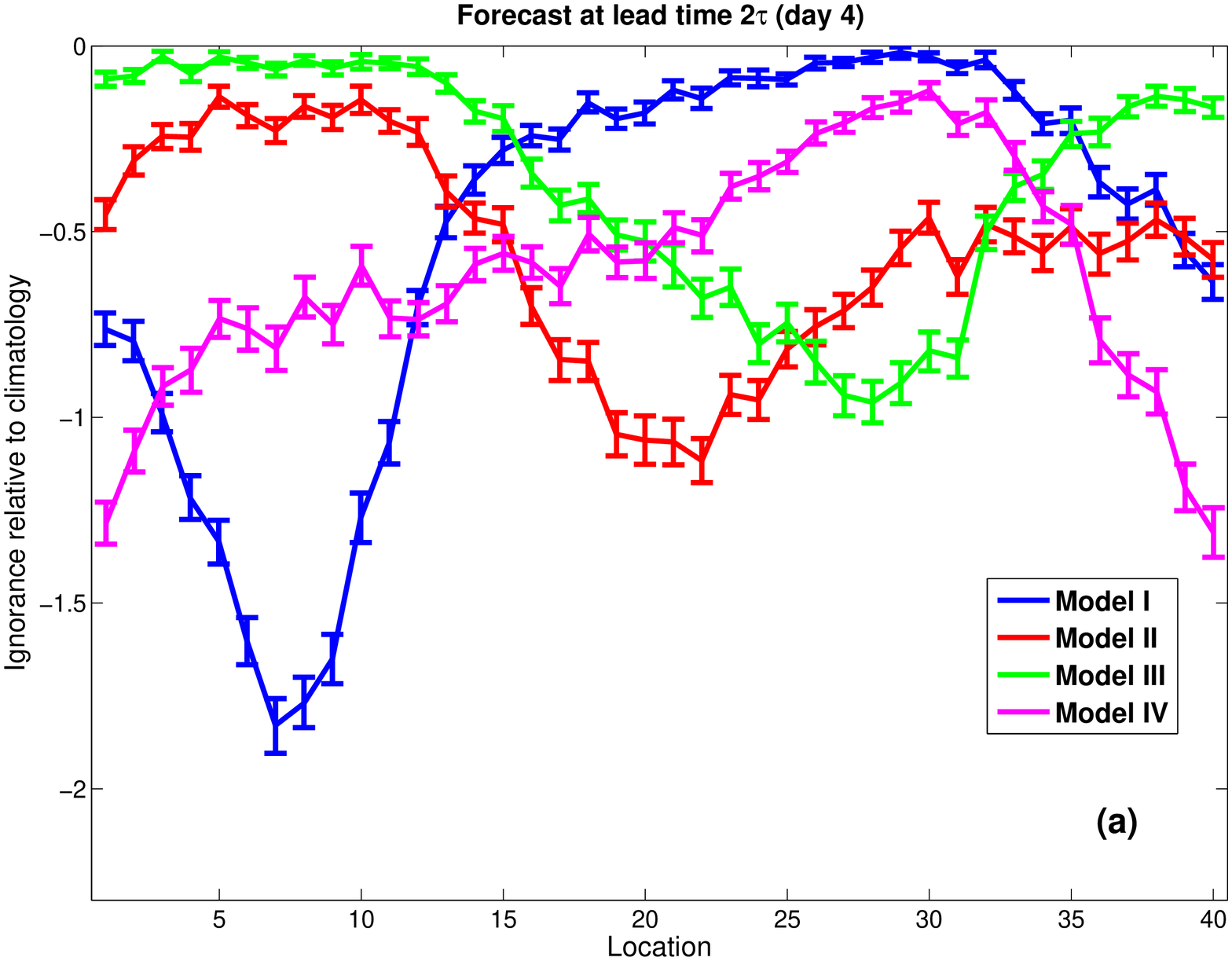, width=0.48\columnwidth, height=6cm}
  \epsfig{file=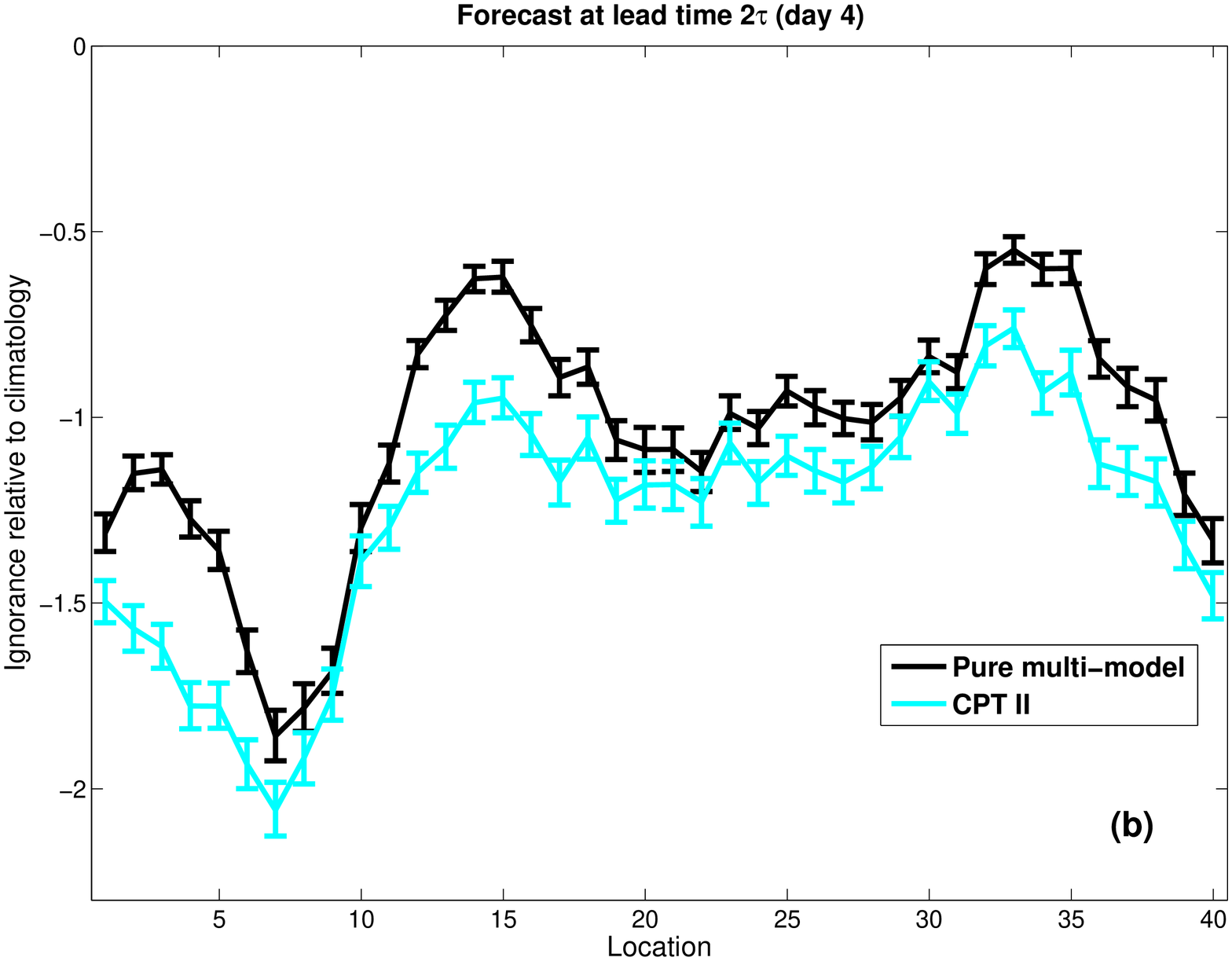, width=0.48\columnwidth, height=6cm}
}

\caption{Ignorance score of forecasts as a function of location (dimension) at lead time $2\tau=0.8$ time unit, a) forecasts from each individual model, b) pure multi-model forecast (Black) and CPT II forecast (Cyan).}
\label{fig:maperr1}
\end{figure}
 
\begin{figure}[!h]
\hbox{
  \epsfig{file=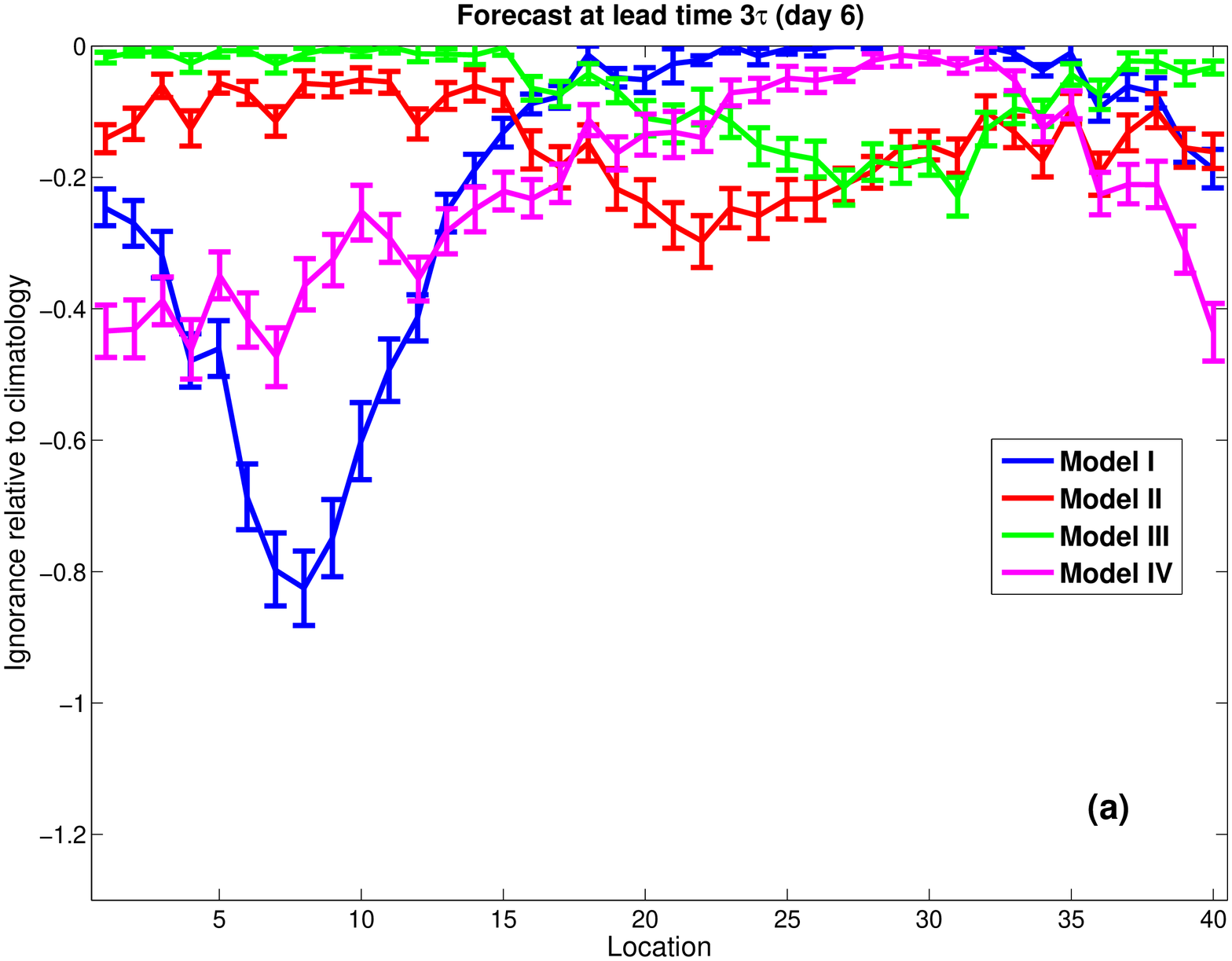, width=0.48\columnwidth, height=6cm}
  \epsfig{file=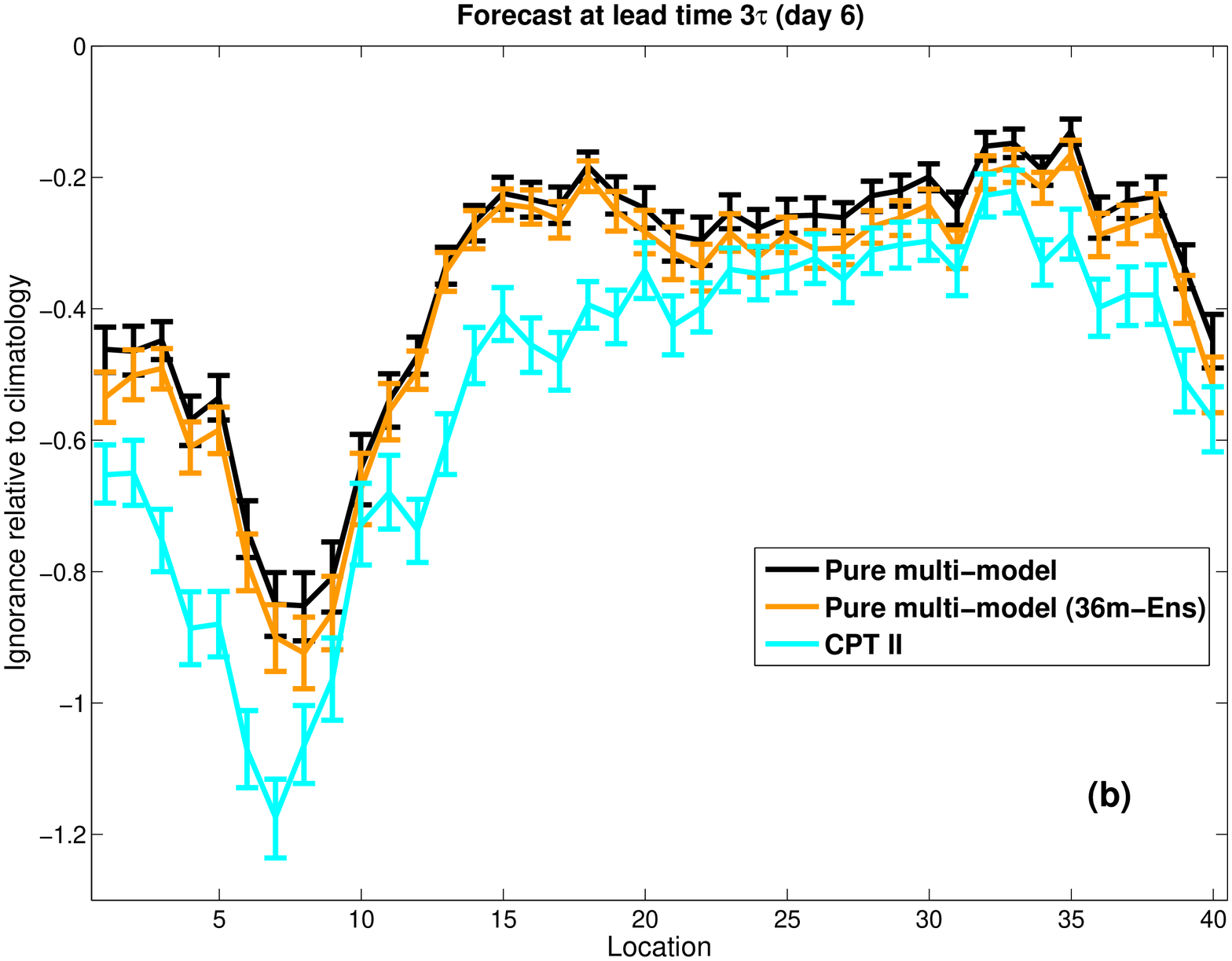, width=0.48\columnwidth, height=6cm}
}

\caption{Ignorance score of forecasts as a function of location (dimension) at lead time $3\tau=1.2$ time unit, a) forecasts from each individual model, b) pure multi-model forecast (Black), pure multi-model forecast with 36-member ensemble from each model (Brown) and CPT II forecast (Cyan).}
\label{fig:maperr1}
\end{figure}

Figure 1a, 2a and 3a shows the empirical Ignorance relative to climatology for each model at different location (dimension) at lead time $\tau$, $2\tau$ and $3\tau$. The empirical Ignorance is calculated based on 2048 forecast outcome pairs and the climatological distribution is estimated using 2048 historical observations.

A simple pruning algorithm is adopted to maintain a manageable ensemble size, which prunes four model (9-member) ensemble forecast trajectories into 9 sequences of states (see Equation (1)): Define the pruned ensemble orbits to be $\{\textbf{Y}(1),\dots,\textbf{Y}(9)\}$ where $\textbf{Y}(j)\equiv\{ \textbf{y}(t_{0};j),\textbf{y}(t_{0}+\Delta t;j),\dots,\textbf{y}(t_{0}+\tau;j) \}$ and $\textbf{y}(t;j)\equiv\{ y_{1}(t;j),y_{2}(t;j),\dots,y_{40}(t;j) \}\in \mathbb{R}^{40}$. Assign the value of $x_{i}(t,B,j)$ to $y_{i}(t;j)$, where $B$ is historically the local best model among (I, II, III, IV) that produce best (Ignorance) forecasts at lead time $\tau$ for dimension location $i$. 

To demonstrate the use of CPT II approach, the outputs from pure model approach are compared with the results from CPT II approach at lead time $\tau$, $2\tau$ and $3\tau$. Note that both the outputs from pure model runs and those from CPT II approach at any lead time $t$ are multi-model ensemble. The multi-model ensemble is interpreted by combining the forecast distributions, generated from ensembles of individual model outputs to produce a single probabilistic multi-model forecast distribution for evaluation. A linearly weighted approach is adopted to combine the single model forecast distribution (for a full description see~\cite{Higgins16}, and Appendix C).

Figure 1b, 2b and 3b compares the probabilistic forecasts from the pure multi-model outputs(Black) with those from CPT II(Cyan) at lead time $\tau$, $2\tau$ and $3\tau$. %The combined forecast is the weighted linear sum of the constituent distributions.   
At lead time $\tau$, the dynamical information from each individual are combined and embedded into the forecasts which are also the initial states of the forecasts for the next cross pollination period. Such additional information in the initial conditions reveals its value at the next forecast period, where significant improvement in skill is shown at lead time $2\tau$ and $3\tau$. Note in figure 3b that the CPT II forecast also significantly outperforms the pure multi-model forecast (Brown) based on four times larger ensemble size (each model produces 36-member ensemble forecast).

{In this paper, CPT II exploits the sophisticated PDA data assimilation scheme~\cite{Du2014a,Du2014b} to allow selective inclusion of locations (state space components) in the forecast simulation of each model. Doing so allows CPT forecast system to general forecasts with support beyond that of with simple model forecast systems or traditional multi-model forecast systems. As demonstrated above, it is possible that this approach increases forecast skill.}

\section{Conclusion}

Suppose for a moment one has two models which simulate supply and demand, given the current supply and demand. Model A produces significantly better forecasts of supply, while Model B yields significantly better forecasts of demand. The traditional multi-model approach is to consider an ensemble of simulation under model A and a second independent ensemble of simulations under model B. The specific model inadequacy in each model will result in a decay in the relevance of the probabilistic forecasts with lead time. Cross Pollination in Time II aims to delay this decay, extending the lead-time of informative forecasts, by assimilating forecasts of the near-term future to generate an enhanced ensemble of forecasts in the medium range, and iterating the process into the long range. In this simple case, taking the ``supply" forecast from model A and the ``demand" forecast from Model B would produce a new initial condition to be folded into the forecast ensembles of each model at lead time one and propagated into the future, improving the ability to forecast.

By contract, the state of a modern weather model consists of (more than) tens of millions of components, and no two operational models actually share the same model-state space. CPT II overcomes this challenge by using a data assimilation designed to start with a pseudo-orbit, and an initial pseudo-orbit extracted from the initial simulation trajectory of each model. This approach yields probabilistic forecasts more skillful than traditional approaches, even when the ensemble size of the traditional approach is increased by a factor of four. As noted above, challenges remain in deploying and interpreting CPT II forecast systems; this concrete example of success is intended to motivate exploration of more realistic cases.

\begin{center}
      {\bf APPENDIX}
    \end{center}

\begin{appendix}
\appendix

\section{Pseudo-orbit Data Assimilation}

A brief description of the PDA approach is given in the following paragraphs (for more details, see Du and Smith~\cite{Du2014a,Du2014b}).

Let the dimension of the model-state space be $m$ and the number of observation times in the assimilation window be $n$ (for experiments presented in this paper, $n=\frac{\tau}{\Delta t}$). A {\it pseudo-orbit}, $\textbf{U}\equiv \{\textbf{u}_{-n+1},...,\textbf{u}_{-1},\textbf{u}_{0}\}$, to is a point in the $m \times n$ dimensional sequence space for which $\textbf{u}_{t+1}\neq F(\textbf{u}_{t})$ for any component of $\textbf{U}$. Define the component of the {\it mismatch} of a pseudo-orbit $\textbf{U}$ at time $t$ to be $\textbf{e}_{t}=\mid F(\textbf{u}_{t})-\textbf{u}_{t+1} \mid$, $t=-n+1,...,-1$ and the mismatch cost function to be
\begin{eqnarray}
 \label{eq:miscost}
    C(\textbf{U})=\sum\textbf{e}_{t}^{2}
\end{eqnarray}
The {\it Pseudo-orbit Data Assimilation} minimizes the mismatch cost function for $\textbf{U}$ in the $m\times n$ dimensional sequence space via gradient descent (GD) minimization algorithm. For CPT II, such minimization is initialized with the combined model output of sequence states in the observation space, $\textbf{Y(j)}$ in Equation (1) and (2). An important advantage of PDA is that the minimization is done in the sequence space: information from across the assimilation window is used simultaneously. 

The pseudo-orbit $\textbf{U}$ is updated on every iteration of the GD minimization. Let the result of the GD minimization be $^{\alpha}\textbf{U}$ where $\alpha$ indicates algorithmic time in GD. Due to the model imperfection, the minimization is applied with a stopping criteria in order to obtain more consistent pseudo-orbits~\cite{Du2014b} as the mismatches reflect the point-wise model error, which is known to exist when the model is imperfect. In the experiments presented in this paper the stopping criteria targeted forecast performance at lead time $\tau$, $2\tau$ and $3\tau$. 

\section{Ensemble Interpretation}

An ensemble of simulations is transformed into a probabilistic distribution function by a combination of kernel dressing and blending with climatological distribution (see \cite{Brocker08}). An $N$-member ensemble at time $t$ is given as $X_{t}=[x^{1}_{t},...,x^{N}_{t}]$, where $x^{i}_{t}$ is the $i$th ensemble member. For simplicity, all ensemble members under a model are treated as exchangeable. Kernel dressing defines the model-based component of the density as:
\begin{eqnarray}
 \label{eq:SKD}
 p(y:X,\sigma)=\frac{1}{N\sigma}\sum^{N}_{i}K\left(\frac{y-(x^{i}-\mu)}{\sigma}\right),
\end{eqnarray}
\noindent where $y$ is a random variable corresponding to the density function $p$ and $K$ is the kernel, taken here to be
\begin{eqnarray}
 \label{eq:SKD1}
  K(\zeta)=\frac{1}{\sqrt{2\pi}}exp(-\frac{1}{2}\zeta^{2}).
\end{eqnarray}
\noindent Thus each ensemble member contributes a Gaussian kernel centred at $x^{i}-\mu$. Here $\mu$ is an offset, which accounts for any systematic ``bias''. For a Gaussian kernel, the kernel width $\sigma$ is simply the standard deviation.

{For any finite ensemble, there remains the chance of $\sim\frac{2}{N}$ that the outcome lies outside the range of the ensemble even when the outcome is selected from the same distribution as the ensemble itself. Given the nonlinearity of the model, such outcomes can be very far outside the range of the ensemble members. In addition to $N$ being finite, in practice, of course, the simulations are not drawn from the same distribution as the outcome as the ensemble simulation system is not perfect. To improve the skill of the probabilistic forecasts, the kernel dressed ensemble may be blended with an estimate of the climatological distribution of the system (see~\cite{Brocker08} for more details, {\cite{Roulston03} for an alternative kernels} and~\cite{Raftery05} for a Bayesian approach).} The blended forecast distribution is then written as
\begin{eqnarray}
 \label{eq:blending}
    p(\cdot)=\alpha p_{m}(\cdot)+(1-\alpha)p_{c}(\cdot),
\end{eqnarray}
\noindent where $p_{m}$ is the density function generated by dressing the model ensemble and $p_{c}$ is the estimate of climatological density. {The blending parameter $\alpha$ determines how much weight is placed in the model. Specifying the three values (the offset $\mu$, the kernel width $\sigma$, and the blended parameter $\alpha$) defines the forecast distribution. }These parameters are fitted simultaneously by optimising the empirical Ignorance score in the training set.

\section{Weighting Multi-model Ensemble}

There are many ways in which forecast distributions, generated from ensembles of individual model runs can be combined to produce a single probabilistic multi-model forecast distribution. One approach may be to assign equal weight to each model and simply sum the distributions generated from each model to obtain a single probabilistic distribution (see \cite{Hagedorn05}). In general, different forecast models do not provide equal amounts of information, one may want to weight the models according to some measure of past performance, see for example~\cite{Rajagopalan02,Doblas05}. The combined multi-model forecast is the weighted linear sum of the constituent distributions,
\begin{eqnarray}
 \label{eq:weight}
    p_{mm}=\sum_{i}\omega_{i}p_{i},
\end{eqnarray}
\noindent where the $p_{i}$ is the forecast distribution from model $i$ and $\omega_{i}$ its weight, with $\sum_{i}\omega_{i}=1$. The weighting parameters may be chosen by minimizing the Ignorance score for example, although fitting $\omega_{i}$ in this way can be costly and is typically complicated by different models sharing information. {And, of course, the weights of individual models are expected to vary as a function of lead time.} 

To avoid ill fitting model weights a simple iterative method to combine models is adopted instead of fitting all the weights simultaneously. For each lead time, the best (in terms of Ignorance) model is first combined with the second best model to form a combined forecast distribution (by assign weights to both models). The combined forecast distribution is then combined with the third best model to update the combined forecast distribution. Repeat this process until the worst model is included. 

\end{appendix}

\section*{Acknowledgment}
This research was supported by the LSE's Grantham Research Institute on Climate Change and the Environment and the ESRC Centre for Climate Change Economics and Policy, funded by the Economic and Social Research Council and Munich Re; it was also funded as part of the EPSRC-funded Blue Green Cities (EP/K013661/1). Additional support for H.D. was also provided by the National Science Foundation Award No. 0951576 ``DMUU: Center for Robust Decision Making on Climate and Energy Policy (RDCEP)". L.A. S. gratefully acknowledges the continuing support of Pembroke College, Oxford.


\begin{thebibliography}{990}
%\begin{thebibliography}
%\bibliographystyle{plain}
%\bibliography{dubib}

\bibitem{Bernardo79}
J. M. Bernardo. Expected information as expected utility. Ann. Stat., 7:686-690, 1979.

\bibitem{Bishop01}
C. H. Bishop, B. J. Etherton, and S. J. Majumdar. Adaptive Sampling with the Ensemble Transform Kalman Filter. Part I: Theoretical Aspects. Mon. Wea. Rev., 129(3):420-436, 2001.

\bibitem{Brocker07}
J. Brocker and L.A. Smith. Scoring probabilistic forecasts: On 356 the importance of being proper. Wea. Forecasting, 22:382-388, 2007.

\bibitem{Brocker08}
J. Brocker and L.A. Smith. From ensemble forecasts to predictive distribution functions. Tellus, 60:663–678, 2008.

\bibitem{Doblas05}
F. J. Doblas-Reyes, R. Hagedorn, and T. N. Palmer. The rationale behind the success of multi-model ensembles in seasonal forecasting. part ii: Calibration and combination. Tellus A, 57:234, 2005.

\bibitem{Du2014a}
H. Du and L.A. Smith. Pseudo-orbit data assimilation part I: the perfect model scenario. Journal of the Atmospheric Sciences, 71(2):469-482, 2014.

\bibitem{Du2014b}
H. Du and L.A. Smith. Pseudo-orbit data assimilation part II: assimilation with imperfect models. Journal of the Atmospheric Sciences, 71(2):483-495, 2014.

\bibitem{Good52}
I. J. Good. Rational decisions. Journal of the Royal Statistical Society, XIV(1), 1952.

\bibitem{Hagedorn05}
R. Hagedorn, F. J. Doblas-Reyes, and T. N. Palmer. The rationale behind the success of multi-model ensembles in seasonal forecasting. part i: Basic concept. Tellus A, 57:219, 2005.

\bibitem{Hagedorn09}
R. Hagedorn and L. A. Smith. Communicating the value of probabilistic forecasts with weather roulette. Meteor. Appl., 16:143155, 2009.

\bibitem{Harrison95}
M. S. J. Harrison, T. N. Palmer, D. S. Richardson, R. Buizza, and T. Petroliagis. Joint ensembles from the ukmo and ecmwf models. In ECMWF Seminar Procedings: Predictability, volume 2, page 61120, ECMWF, Reading, UK, 1995.

\bibitem{Higgins16}
S. Higgins, H. Du, and L.A. Smith. On the design and use of ensembles of multi-model simulations for forecasting. In preparation for Nonlinear Processes in Geophysics, 2016.

\bibitem{Judd04}
K. Judd and L.A. Smith. Indistinguishable states ii: The imperfect model scenario. Physica D, 196:224, 2004.

\bibitem{Kirtman13}
B. P. Kirtman, D. Min, J. M. Infanti, J. L. Kinter, D. A. Paolino, Q. Zhang, H. van den Dool, S. Saha, M. P. Mendez, E. Becker, P. Peng, P. Tripp, J. Huang, D. G. DeWitt, M. K. Tippett, A. G. Barnston, S. Li, A. Rosati, S. D. Schubert, M. Rienecker, M. Suarez, Z. E. Li, J. Marshak, Y. Lim, J. Tribbia, K. Pegion, W. J. Merryfield, B. Denis, and E. F. Wood. The North American Multimodel Ensemble: Phase-1 Seasonal-to-Interannual Prediction; Phase-2 toward Developing Intraseasonal Prediction. Bull. Amer. Meteor. Soc., 95(4):585-601, 2013.

\bibitem{Lorenz96}
E. N. Lorenz. Predictability - a problem partly solved. Cambridge University Press, 1996.

\bibitem{Palmer04}
T. N. Palmer, F. J. Doblas-Reyes, R. Hagedorn, A. Alessandri, S. Gualdi, U. Andersen, H. Feddersen, P. Cantelaube, J. M. Terres, M. Davey, R. Graham, P. D´el´ecluse, A. Lazar, M. D´equ´e, J. F. Gu´er´emy, E. D´ıez, B. Orfila, M. Hoshen, A. P. Morse, N. Keenlyside, M. Latif, E. Maisonnave, P. Rogel, V. Marletto, and M. C. Thomson. Development of a european multimodel ensemble system for seasonal-to-interannual prediction (demeter). Bull. Amer. Meteor. Soc., 85(6):853-872, 2004.

\bibitem{Raftery05}
A. E. Raftery, T. Gneiting, F. Balabdaoui, and M. Polakowski. Using bayesian model averaging to calibrate forecast ensembles. Mon. Wea. Rev., 133:1155-1174, 2005.

\bibitem{Rajagopalan02}
B. Rajagopalan, U. Lall, and S. E. Zebiak. Categorical climate forecasts through regularization and optimal combination of multiple gcm ensembles. Monthly Weather Review,130:1792-1811, 2002.

\bibitem{Roulston02}
M. S. Roulston and L. A. Smith. Evaluating probabilistic forecasts using information theory. Mon. Wea. Rev., 130:1653-1660, 2002.

\bibitem{Roulston03}
M. S. Roulston and L. A. Smith. Combining dynamical and statistical ensembles. Tellus, 55:16-30, 2003.

\bibitem{Smith00}
L. A. Smith. Disentangling uncertainty and error: On the predictability of nonlinear systems. In Alistair I. Mees, editor, Nonlinear Dynamics and Statistics, chapter 2, pages 31-64. Birkha¨user Boston, 2000.

\bibitem{Smith14}
L. A. Smith, Hailiang Du, Emma B. Suckling, and Falk Niehrster. Probabilistic skill in ensemble seasonal forecasts. Quarterly Journal of the Royal Meteorological Society, 141(689):1085-1100, 2015.

\bibitem{Smith02}
L.A. Smith. What might we learn from climate forecasts? In Proc. National Acad. Sci., volume 4, pages 2487-2492, USA, 2002.

\bibitem{Smith15}
L.A. Smith, E.B. Suckling, E.L. Thompson, T. Maynard, and H. Du. Towards improving the framework for probabilistic forecast evaluation. Climatic Change, 2015.

\bibitem{Stephenson05}
D.B. Stephenson, C. Coelho, F. Doblas-Reyes, and M. Alonso Balmaseda. Forecast assimilation: a unified framework for the combination of multimodel weather and climate predictions. Tellus A, 57:253-264, 2005.

\bibitem{Wang09}
B. Wang, J. Lee, I. Kang, J. Shukla, C. K. Park, A. Kumar, J. Schemm, S. Cocke, J. S. Kug, J. J. Luo, T. Zhou, B. Wang, X. Fu, W. T. Yun, O. Alves, E. Jin, J. Kinter, B. Kirtman, T. Krishnamurti, N. Lau, W. Lau, P. Liu, P. Pegion, T. Rosati, S. Schubert, W. Stern, M. Suarez, and T. Yamagata. Advance and prospectus of seasonal prediction: assessment of the APCC/CliPAS 14-model ensemble retrospective seasonal prediction (1980-2004). Climate Dynamics, 33(1):93-117, 2009.

\bibitem{Weisheimer09}
A. Weisheimer, F. J. Doblas-Reyes, T. N. Palmer, A. Alessandri, A. Arribas, M. D´equ´e, N. Keenlyside, M. MacVean, A. Navarra, and P. Rogel. ENSEMBLES: A new multi-model ensemble for seasonal-to-annual predictions skill and progress beyond DEMETER in forecasting tropical Pacific SSTs. Geophysical Research Letters, 36(21), 2009.

\bibitem{Wilks06}
D. S. Wilks. Comparison of ensemble-MOS methods in the Lorenz’96 setting. Meteorological Applications, 13:243-256, 2006.

\bibitem{Wilks07}
D. S. Wilks and T. M. Hamill. Comparison of Ensemble-MOS Methods Using GFS Reforecasts. Mon. Wea. Rev., 135(6):2379, 2007.

\bibitem{Wilks05}
D.S. Wilks. Statistical Methods in the Atmospheric Sciences. Academic Press, second edition, 2005.

% \bibitem{mcsharrys99}
% P.E. McSharry and L.A. Smith, Better nonlinear models from noisy data: Attractors with maximum likelihood, Phys. Rev. Lett {\bf 83}, (21): 4285-4288, (1999).

% \bibitem{proper}
% J. M. Bernardo. Expected information as expected utility. Annals of Statistics, 7(7):686�C690, (1979).

% \bibitem{Lorenz65}
% E.N. Lorenz, A study of the predictability of a 28-variable atmospheric model. Tellus, 17, 321-333, (1965)

% \bibitem{IS3}
% L.A. Smith, M.C. Cuellar, H. Du and K. Judd, Exploiting Dynamical Coherence: A Geometric Approach to Parameter Estimation in Nonlinear Models, submitted to {\it Phys. Lett. A}, (2008)

% \bibitem{IS1}
% K. Judd and L.A. Smith, Indistinguishable States I: The Perfect Model Scenario, {\it Physica D} {\bf 151}: 125-141, (2001).

% \bibitem{Account95}
% L. Smith, Accountability and error in ensemble forecasting, in Predictability, seminar Proceedings, pp. 351-369, ECMWF, Shinfield Park, Reading, Berkshire, RG29ax, (1995).

% \bibitem{IgnRS}
% M.S. Roulston and L.A. Smith, Evaluating probabilistic forecasts using information theory, {\it Monthly Weather Review}, {\bf 130}, 1653-1660, (2002)

% \bibitem{EnKF94}
% G. Evensen, Sequential data assimilation with a nonlinear quasi-geostrophic model using Monte Carlo methods to forecast error statistics, {\it J. Geophys. Res.}, {\bf 127}, 2128-2142, (1994).

% \bibitem{Adress}
% J. Brocker and L.A. Smith, From ensemble forecasts to predictive distribution functions, {\it Tellus A}, 60, 663-678 (2007)

% \bibitem{Score}
% J. Brocker, L.A Smith, Scoring Probabilistic Forecasts: On the Importance of Being Proper, {\it Weather and Forecasting}, 22 (2), 382-388, (2006)

% \bibitem{Wilks06}
% D.S. Wilks, Comparison of ensemble-MOS methods in the Lorenz'96 setting. {\it Meteorological Applications}, 13, (2006).

% \bibitem{Good}
% G.W. Brier, Verification of forecasts expressed in terms of probabilities. {\it Mon. Wea. Rev.}, {\bf 78}, 1�C3, (1950).



% \bibitem{IS2}
% K. Judd and L.A. Smith, Indistinguishable States II: The Imperfect Model Scenario, {\it Physica D} {\bf 196}: 224-242, (2004).

% \bibitem{GDDav}
% M.E.Davies, Noise reduction by gradient descent, Int. J. Bifurcation Chaos 3, 113-118, (1992).

% \bibitem{GDJud}
% D.Ridout and K.Judd, Convergence properties of gradient descent noise reduction. {\it Physica D} {\bf 165} 27-48, (2001).

%  \bibitem{press92}
% W.~H. Press, B.~Flannery, S.~Teukolsky, and W.~Vetterling, {\em
%   {N}umerical {R}ecipes in {C}} (CUP, Cambridge, 1992).

% \bibitem{GFD}
% K. Judd and L.A. Smith and A. Weisheimer,
% Gradient free descent: shadowing and state estimation with limited derivative information, {\it Physica D}, 190, 153--166 (2004).

% \bibitem{ishadows1} L.A. Smith, Disentangling Uncertainty and
% Error: On the Predictability of Nonlinear Systems. Nonlinear Dynamics and Statistics, A.I.Mees, Ed., Birkhauser, 31-64 (2000).

% \bibitem{ishadows2}
% I. Gilmour, Nonlinear model evaluation: $\iota$-shadowing, probabilistic prediction and weather forecasting, Ph.D. thesis, University of Oxford, (1998).

% \bibitem{Jim98}

% J.A.Hansen, Adaptive observations in spatialy-extended nonlinear dynamical systems, Ph.D. thesis, University of Oxford, (1998).

% \bibitem{Pissor04}
% V.F. Pisarenko and D. Sornette, Statistical methods of parameter estimation for deterministically chaotic time series, Phys Rev E, {\bf 69} (2004).

% \bibitem{Timmer98}
% J. Timmer, Modeling noisy time series: physiological tremor, Int. J. Bif. Chaos {\bf 8}, 1505�C1516 (1998).

% \bibitem{judd_bayes}
% K. Judd, Chaotic-time-series reconstruction by the {B}ayesian paradigm: {R}ight results by wrong methods, Physics Review Letters, {\bf 67}, {026212},(2003).

% \bibitem{NRedux1}
%  M. Casdagli, S. Eubank, J.~D. Farmer, and J. Gibson, State space reconstruction in the presence of noise, Physica D,
% {\bf 51},  52 (1991);

% \bibitem{NRedux2}
% C.L.Bremer and D.T.Kaplan, Markov chain Monte Carlo estimation of nonlinear dynamics from time series, Physica D {\bf 160}
% 116--126 (2001).

% \bibitem{NRedux3}

%  P. Grassberger, T. Schreiber, and C. Schaffrath, Non--linear time sequence analysis, Int. J. Bif. and
% Chaos {\bf 1},  521  (1991).

% \bibitem{henon76}
% M. H\'{e}non, A two-dimensional mapping with a strange attractor, Commun. Math. Phys. {\bf 50}  69-77  (1976).

% \bibitem{ikeda} %Ikeda (jones et al)
% K.Ikeda, Multiple valued stationarity state and its instability of the transmitted light by a ring cavity system, Optical Communications, {\bf 30} 257-261 (1979).

% \bibitem{oldshadows1}
% E.J.  Kostelich and J.A. Yorke, Noise Reduction in Dynamical Systems, Phys Rev A 38, 1649, (1988).

% \bibitem{oldshadows2}
% T. Sauer, J.A. Yorke, M. Casdagli, Embedology, J. Stat. Phys. 65, 579, (1991).

% \bibitem{borel}
% E. Borel, {\it Probability and Certainty} Walker, New York, (1963).

% \bibitem{berliner91}
% M.L. Berliner, Likelihood and Bayesian Prediction for Chaotic Systems, J. Am. Stat. Assoc. {\bf 86},  938-952  (1991).

% \bibitem{newhouse}
% Guckenheimer, J. Moser and S. Newhouse, Dynamical Systems, CIME Lectures, Birkhauser, 289pp., (1980).

% \bibitem{stainforth}
% D.A.Stainforth, et al. Uncertainty in predictions of the climate response to rising levels of greenhouse gases, Nature, 433, 403�C406, (2005).

% \bibitem{goodbayes}
% D.S. Sivia, Data analysis - a Bayesian tutorial, Clarendon Press, Oxford University Press, (1997).

% %\bibitem{goodbayes1}
% %Berger, J.P.M. Heald, J. Stark (preprint)

% \bibitem{jaeger96a}
% L. Jaeger and H. Kantz, Unbiased reconstruction of the dynamics underlying a noisy chaotic time series, Chaos {\bf 6},  440  (1996).

% \bibitem{jaeger96b}
% H. Kantz and L. Jaeger, Homoclinic tangencies and non-normal Jacobians - effects of noise in non-hyperbolic systems, Physica D {\bf 105},  79-96  (1997).

% \bibitem{kostelich92}
% E.J. Kostelich, Problems in estimating dynamics from data, Physica D {\bf 58},  138  (1992).

% \bibitem{MS66}
% W. D.Moore and E. A.Spiegel, A thermally excited nonlinear oscillator. The Astrophysical Journal. Volume 143, pp. 871-887, (1966).

% \bibitem{loz96}
% E.N. Lorenz, Predictability A problem partly solved, in: ECMWF Seminar Proceedings on Predictability, Reading, United Kingdom, ECMWF, (1995).

% \bibitem{ZiehmannSK}
% C. Ziehmann, L.A Smith and J. Kurths, Localized Lyapunov Exponents and the Prediction of Predictability, Phys. Lett. A, 271 (4): 237-251, (2000).




%\bibitem{casella90}
%G. Casella and R.~L. Berger, {\em Statistical Inference}
%(Wadsworth \&
%  Brooks/Cole, California, 1990).


\end{thebibliography}
\end{document}